\documentclass[12pt,preprint]{aastex}
\usepackage{amsmath}

\shorttitle{Activated asteroid P/2013 P5 (PANSTARRS)}
\shortauthors{Moreno et al.}

\begin{document}

%% LaTeX will automatically break titles if they run longer than
%% one line. However, you may use \\ to force a line break if
%% you desire.

\title{Intermittent dust mass loss from activated asteroid
  P/2013 P5 (PANSTARRS)}

%% Use \author, \affil, and the \and command to format
%% author and affiliation information.
%% Note that \email has replaced the old \authoremail command
%% from AASTeX v4.0. You can use \email to mark an email address
%% anywhere in the paper, not just in the front matter.
%% As in the title, use \\ to force line breaks.

\author{F. Moreno\affil{Instituto de Astrof\'\i sica de Andaluc\'\i a, CSIC,
  Glorieta de la Astronom\'\i a s/n, 18008 Granada, Spain}
\email{fernando@iaa.es}}

\author{
J. Licandro\affil{Instituto de Astrof\'\i sica de Canarias,
  c/V\'{\i}a 
L\'actea s/n, 38200 La Laguna, Tenerife, Spain, 
\and 
 Departamento de Astrof\'{\i}sica, Universidad de
  La Laguna (ULL), E-38205 La Laguna, Tenerife, Spain}}

\author{
C. \'Alvarez-Iglesias\affil{Instituto de Astrof\'\i sica de Canarias,
  c/V\'{\i}a 
L\'actea s/n, 38200 La Laguna, Tenerife, Spain, 
\and 
 Departamento de Astrof\'{\i}sica, Universidad de
  La Laguna (ULL), E-38205 La Laguna, Tenerife, Spain, 
\and 
GTC Project, E-38205 La Laguna, Tenerife, Spain}}

\author{
A. Cabrera-Lavers\affil{Instituto de Astrof\'\i sica de Canarias,
  c/V\'{\i}a 
L\'actea s/n, 38200 La Laguna, Tenerife, Spain, 
\and 
 Departamento de Astrof\'{\i}sica, Universidad de
  La Laguna (ULL), E-38205 La Laguna, Tenerife, Spain, 
\and 
GTC Project, E-38205 La Laguna, Tenerife, Spain}}  

\and

\author{
F. Pozuelos\affil{Instituto de Astrof\'\i sica de Andaluc\'\i a, CSIC,
  Glorieta de la Astronom\'\i a s/n, 18008 Granada, Spain}}

%% Notice that each of these authors has alternate affiliations, which
%% are identified by the \altaffilmark after each name.  Specify alternate
%% affiliation information with \altaffiltext, with one command per each
%% affiliation.

\begin{abstract}

We present observations and models of the dust environment of
activated asteroid P/2013 P5 (PANSTARRS). The object displayed a
complex morphology during the observations, with the presence of
multiple tails. We combined our own observations, all made with
instrumentation attached to the
10.4m Gran Telescopio Canarias (GTC) on La Palma, with previously
published Hubble Space Telescope (HST) images to build a model aimed
at fitting all 
the observations. Altogether, the data cover a full 3-month period of
observations which can be explained by an intermittent dust loss. The
most plausible scenario is that of an asteroid rotating with the spinning
axis oriented perpendicular to the orbit plane and loosing mass from
the equatorial region, consistent with a rotational break-up.  
Assuming that the ejection velocity of
the particles ($v \sim$0.02-0.05 m s$^{-1}$) 
corresponds to the escape velocity, the object diameter is constrained to
 $\sim$30-130 m for bulk densities 3000 to 1000 kg m$^{-3}$.

\end{abstract}

\keywords{Minor planets, asteroids: individual (P/2013 P5 (PANSTARRS) --- 
Methods: numerical}

\section{Introduction}

Activated asteroid P/2013 P5 (PANSTARRS) was discovered by 
Pan-STARRS survey as a 21st magnitude comet on August 15.50, 2013
\citep{Micheli13}. This object has a typical inner-belt asteroid orbit
and yet displays a cometary-like tail, so it can be classified as a
Main-belt comet (MBC). The object shares similar
orbital elements with previously discovered disrupted asteroid P/2010
A2 (LINEAR) \citep{Jewitt10,Snodgrass10,Moreno13}, both 
belonging to the Flora collisional family. The origin of activity
taking place in the MBCs is unknown for most of the those objects. While
some have been associated to impulsive events, such as collisions
with another body or rotational break-up, other are most likely
linked to water-ice sublimation. For reviews on those objects and the
likely mechanisms involved in their activity, see
e.g. \cite{Hsieh06}, \cite{Bertini11}, and \cite{Jewitt12}.

A series of stunning images from the HST \citep{Jewitt13} at two
epochs reveal the asteroid as a multiple-tailed object. Using the
Finson-Probstein formalism, these tails
have been associated to a series of ejection events at different
dates, and the likely cause of the activity has been linked to a
rotational disruption. In this paper, we present our own data,
that were acquired during two months after the HST observations, and
combine them with the HST data. Our
aim is then to monitor the activity scenario during a longer time
frame and, mostly, to characterize the dust activity in terms of time
variation of the mass loss, particle size distribution, and ejection
velocities. Models mimicking an equatorial mass loss from the object
have been incorporated in an attempt to investigate if a rotational
disruption could be compatible with the ejection scenario.             

\section{Observations and data reduction}

Images through Sloan $r^\prime$ and $g^\prime$ filters 
of P/2103 P5 were recorded under
photometric and excellent seeing conditions (0.8-0.9$\arcsec$) on the nights
of 2013 October 7 and 2013 November 8 (only $r^\prime$ images). We
used the OSIRIS    
Optical System for Image and Low Resolution Integrated Spectroscopy
(OSIRIS) camera-spectrograph \citep{Cepa00,Cepa10} at the GTC. The
OSIRIS instrument consists of two Marconi CCD detectors, each with
2048$\times$4096 pixels and a total unvignetted field of view of
7.8$\arcmin\times$7.8$\arcmin$. The plate scale was 0.127
$\arcsec$ px$^{-1}$, but we used a 2$\times$2 pixel binning in order to
improve the signal-to-noise ratio, so that the spatial resolution of the
images becomes 222 km px$^{-1}$ and 270 km px$^{-1}$ at the observation
dates. The images were 
bias and flat-field corrected using standard techniques, and
calibrated in flux using standard stars. A sequence of five images per
filter were obtained. An average image was then 
obtained from the available images by shifting and stacking the frames
with respect to a reference frame by taking into account the
object's sky motion. We estimate that as a result of both the flux
calibration and the stacking procedure, the total flux uncertainty
in the combined images is $\sim$0.1 mag. The final combined images are
shown in Figure 1. The log of the observations is shown in Table 1. In
that table, the apparent ($m$) and absolute ($H$) magnitudes of a region of 10
pixel aperture radius (2.5$\arcsec$ diameter) centered on the
asteroid optocenter of each image is given. The absolute magnitude is
given as $H=m-2.5\log(\Delta r_h)-\Phi(\alpha)$, where
$\Delta$ and $r_h$ are the geocentric and heliocentric distances of
the asteroid, and $\Phi(\alpha)$
is the phase function, which is assumed to be that of an S-type
asteroid, as most objects in the inner asteroid belt. The quantity
$\Phi(\alpha)$ is computed by the \cite{Bowell89} formalism, using a
phase function parameter $g$=0.25, which is typical of S-type
asteroids, the most common objects in the inner belt. The phase terms  
become $\Phi(17.7^\circ)$=--0.81 on October 7, and
$\Phi(27.0^\circ)$=--1.07 on November 8, and the absolute magnitude
$H_{r^\prime}$ converges to the value $H_{r^\prime}$=18.0$\pm$0.1 in
both dates. To compare this value to the reported $H_V$ by
\cite{Jewitt13} ($H_V$=18.69 on Sep. 10, and $H_V$=18.54 on Sep. 23),
we need a transformation from $r^\prime$ to $V$ magnitudes.  Using the 
transformation equations of \cite{Fukugita96}, and 
the magnitude of the Sun in the standard Johnson-Cousins filter 
 \citep[$V_\sun$=--26.75,][]{Cox00}, we derive $r^\prime_\sun$=--26.96. If
the object follows a spectral dependence on wavelength similar to that
of the Sun within the $V$ and $r^\prime$ bandpasses (neutral color), the
$V$ magnitudes can be obtained from the $r^\prime$ magnitudes by
adding 0.21 magnitudes. Then, we would get $H_V$=18.2. This would
indicate a brightness increase since Sep. 10 of $\sim$0.5
mag. This result immediately excludes a single event mechanism of dust
ejection from the asteroid, as such event would have resulted in a
brightness decrease as a function of time. Assuming that the
brightness is entirely due to the asteroid, and not to the surrounding
dust, this would led to an asteroid diameter of $\sim$590 m for a
geometric albedo of $p_v$=0.29, typical of Flora family asteroids
\citep{Masiero13}. We will, however, 
assume that such brightness is entirely associated to the
presence of dust surrounding the asteroid, and not to the presence of
a nucleus, its contribution to the brightness being negligible. A
justification of this hypothesis is given in the Results
section on the basis of the low ejection velocities of the dust
particles that we found in the modeling procedure.

\section{The Model}

To perform the analysis of the images, we used our direct Monte Carlo
dust tail model, 
as described in previous works \citep[e.g.,][]{Moreno12a,Fulle10}. 
In that model, we 
compute the trajectory of a large number of particles after being
ejected from an asteroidal 
or cometary surface. Those particles move under the sole influence of
the solar gravity and 
radiation pressure forces, describing a Keplerian orbit around the
Sun. The orbital elements 
of each ejected particle are functions of the ejection velocity and
the $\beta$ parameter \citep[e.g.][]{Fulle89}. This parameter 
can be written as  $\beta =
C_{pr}Q_{pr}/(2\rho r)$, where $C_{pr}$=1.19$\times$ 10$^{-3}$ kg
m$^{-2}$, $Q_{pr}$ is the radiation pressure coefficient, 
and $\rho$ is the particle density. The position of 
each particle in the plane of sky is then computed according to its
orbital elements, and its 
contribution to the tail or coma brightness is evaluated, as a
function of its size and geometric 
albedo. Owing to the many input models, we
are forced to set some of them to a specific value. Then, the
particles are considered spherical, their density is assumed at
$\rho_p$=1000 kg m$^{-3}$, and their refractive index is set at
$1.88+0.71i$, which is typical of carbonaceous 
composition \citep{Edoh83}. Using
Mie theory, we find that the geometric albedo is $p_v\sim$ 0.04,
and that the radiation pressure
coefficient is $Q_{pr}\sim$ 1 for particles of radius 
$r \gtrsim$1 $\mu$m \citep[][their Figure 5]{Moreno12a}. These choices of
density and geometric albedo are highly arbitrary, since we do not know
their real values, and were made actually to facilitate 
comparison with other MBCs analyzed, for which we assumed such values
\citep[e.g.,][]{Moreno10, Moreno13}.

We start by assuming an asteroid nucleus
which is loosing mass from its equator, where centrifugal acceleration
is maximum, uniformly in longitude. This would correspond to a 
mass loss scenario driven by a rotational disruption, as 
suggested by \cite{Jewitt13}. This introduces three more model parameters
to characterize the rotation properties: the 
orientation of the spinning axis with respect to the orbit plane, which is
given by the obliquity, $I$, and the argument of the subsolar meridian
at perihelion, $\Phi$, and the rotational period, $P$ (simple rotation
is assumed). The nucleus is presumably very small, so that the rotation
period should be very short, of the order of $P\lessapprox$3 h   
\citep{Pravec2002}. We assume $P$=3 
h. The exact value of $P$ does not influence the results if the
tail age is much longer than that, as can be anticipated from the
analysis by \cite{Jewitt13}. The rotation parameters $I$ and $\Phi$
are set initially to $I$=0$^\circ$, and $\Phi$=0$^\circ$.  
To simplify, we also  
set all the possible time-variable parameters (except the dust mass
loss rate) to a constant 
value. Thus, the size distribution power index is set to
$\alpha$=--3.5, and the minimum and maximum particle sizes to 50 $\mu$m and 30
cm, respectively. These values were set after extensive experimentation with the
code. Regarding velocities, we employed a function of the kind
$v(\beta)=v_0\beta^{\gamma}$, were we adopt $\gamma$=1/8, i.e., a very
weak dependence of $v(\beta)$ on $\beta$ consistent
with \cite{Moreno12b} in their analysis of disrupted asteroid P/2012 F5
(Gibbs).  The parameter $v_0$ and the dust mass loss rate as a function
of the heliocentric distance are the fitting parameters.  

\section{Results}

The times of significant dust ejection are first estimated from the
best fitting synchrones to the dust tails. This procedure was applied
to the HST images first, owing to their superb spatial resolution, and
then to the GTC images. 
In 
the GTC images, the tails named A to F in \cite{Jewitt13} (see Figure 2), the
oldest being A, are sometimes blended because of poorer spatial
resolution. Thus, in the GTC image of October 7, 2013, we have A, C/B, D,
and E/F (see Figure 1). In addition, a younger tail not seen in the
HST images (named G) appears. On the other hand, the last GTC image of 
November 8, 2013, does not
show the complexity of the others, displaying a single and narrow tail
extending to the northeast (see Figure 1). This is surely connected to
the fact that the angle between Earth and the asteroid orbital plane
($\delta$) is smaller than at the other dates (see Table 1).

The procedure was then to try different mass loss rates at those
times, and set different ejection speeds (distinct $v_0$) until a good fit to
the whole dataset (HST+GTC) is found in terms of dust tail
brightnesses. The synthetic images corresponding to the GTC data are
convolved with a point spread Gaussian function in order to take into
account the seeing conditions during the observations. During the
fitting procedure, we realized that to fit the length of tail ``G'' in
the GTC 2013 October 7 image, we needed to set $r_{min}$=10 $\mu$m at
the time of its peak emission, this being the only modification to the
particle sizes in the time interval of ejection.

The results of the fits to the HST and GTC images are shown in
Figures 2 (left panels) and 3. The model reproduces accurately all the features
present in the HST and GTC images, in terms of brightness, length, and
width. The dust loss rate profile corresponding to those fits is
displayed in Figure 4, resulting in a total dust mass loss of 10$^7$ kg. 
The best fitted ejection velocity is given 
by $v=0.12\beta^{1/8}$ m s$^{-1}$. This corresponds to ejection velocities ranging
from about 0.02 m s$^{-1}$ to 0.07  m s$^{-1}$, for 30 cm to 50 $\mu$m
particles. We have also attempted to fit the images using a constant
value for the ejection velocity for all the particles. We found very similar results to
those of figures 2 (left panels) and 3 when a constant ejection
velocity in the range 
0.02 to 0.05 m s$^{-1}$ is assumed. Regarding the maximum particle
size ejected, we have verified that models having $r_{max}\gtrapprox$1 cm are
compatible with the observations, provided the total mass ejected is
modified accordingly. Thus, if $r_{max}$ is set to its lowest
acceptable limit, $r_{max}$=1 cm, the dust mass loss rate 
would be a factor of $\sim$5 smaller than that shown in
figure 4, i.e., the total dust mass loss would become 2$\times$10$^6$
kg. This constitutes the lower limit of ejected mass, for the assumed
particle density of 1000 kg m$^{-3}$, and geometric albedo $p_v$=0.04.

The range of possible ejection velocities is 0.02-0.05 m
s$^{-1}$. If these values are associated to escape velocities, this
translates to possible asteroid diameters (assumed spherical) 
in the range 30 to 134 m, and masses in
the range 4.6$\times$10$^7$ to 1.3$\times$10$^9$ kg, for assumed
bulk densities of 1000 to 3000 kg m$^{-3}$.   
This size estimate is well below the upper limit of
480$\pm$80 m diameter derived by \cite{Jewitt13} on the basis of magnitude 
measurements of the central
condensation. As those authors recognize, this is an upper limit as
the measurements could include near nucleus dust. We believe that it
is indeed the case, in such a way that the magnitude of the central
condensation is in fact attributable mainly to the dust around the
nucleus, and not to the nucleus itself, whose contribution must be
minimal according to the small size imposed by the escape
velocity. 

Concerning the rotational parameters of the asteroid, we started, as
mentioned, from a scenario in which the rotating axis is
perpendicular to the orbit plane ($I$=0$^\circ$). We have generated
synthetic images by varying both $I$, and $\Phi$, in the full ranges,
0-180$^\circ$, and 0-360$^\circ$, respectively. We found that the only
possible fits correspond to obliquities of either $I\sim$0$^\circ$ or,
$I\sim$180$^\circ$, independently of $\Phi$, i.e., with the
rotating axis nearly perpendicular to the orbit plane, either pointing
to the North or the South of the plane (prograde or retrograde
motion). When the value of $I$ departs significantly from either
0$^\circ$ or 180$^\circ$, tails wider than observed are obtained.   
   
Finally, we have also attempted to reproduce the observed brightness pattern
using an isotropic ejection model, and the results we obtained for the
GTC images are quite similar to those obtained with the above (anisotropic)
model. However,  the HST data are not well reproduced with
this isotropic ejection model, as the oldest tails, especially ``B'',
``C'', and ``A''  become significantly broader than observed. This is
clearly shown in Figure 2, right panels.

\section{Conclusions}  

From the Monte Carlo dust tail modeling of the observations of
activated P/2013 P5 (PANSTARRS) we can extract the following conclusions:  

1) The object has been subjected to an intermittent dust mass loss,
most likely associated to a rotational disruption. This is confirmed
from the analysis of both HST and GTC images. 
The total dust mass released was of the order of  
10$^7$ kg, for particle density of 1000 kg m$^{-3}$ and geometric
albedo $p_v$=0.04.

2) The model of rotational disruption, based on simulations of an 
object that loose mass from its equatorial region, and whose
rotational axis is perpendicular to its orbit plane, reproduces to
the last detail the observed complex brightness pattern at four
different epochs of HST and GTC observations. For obliquities
different from 0$^\circ$ or 180$^\circ$, the fits get much worse.   
On the other hand, an isotropic ejection model does not fit the HST data,
because it produces much more
diffuse tails than observed. 

3) The ejection velocities are very low, of the order of 0.02-0.05 m
s$^{-1}$. This places a limit to the 
size of the object as to be in the  
range 30-134 m for assumed densities of 3000 to 1000 kg m$^{-3}$.

\acknowledgments

This article is based on observations made with the Gran Telescopio
Canarias (GTC), installed in the Spanish Observatorio del Roque de los
Muchachos of the Instituto de Astrof\'\i sica de Canarias, in the island 
of La Palma. 

This work was supported by contracts  AYA2011-30613-C02-01, AYA2012-39691-C02-01, and 
FQM-4555 (Proyecto de Excelencia, Junta de Andaluc\'\i a). 
J. Licandro gratefully acknowledges support from the Spanish ``Ministerio de
Ciencia e Innovaci\'on'' project AYA2012-39115-C03-03.

\clearpage

\begin{figure}[ht]
\centerline{\includegraphics[scale=0.8,angle=-90]{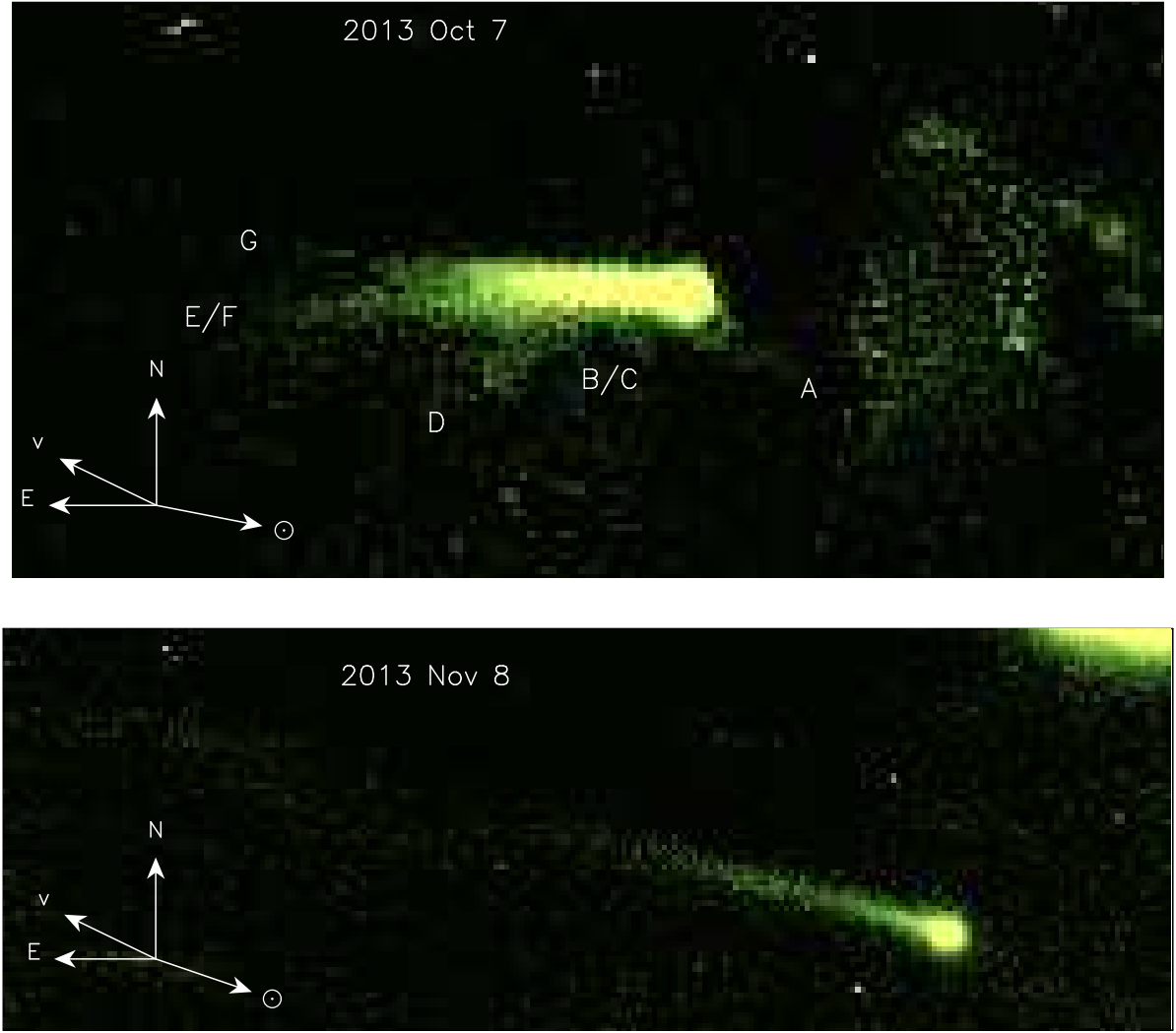}}
\caption{Stacked r' Sloan OSIRIS@GTC images of P/2013 P5 on the
  nights of UT 2013 October 7 (upper panel) and 2013 November 8 (lower
  panel). The dimensions of the upper and lower images are 35488$\times$17744 km
 and 61939$\times$21275 km, respectively. In the upper panel the different tails are marked. The
  nomenclature follows that of \cite{Jewitt13}. Tails marked as B/C and E/F
  are actually a blend of tails B and C, and E and F, respectively, as
  displayed in figure 1 of \cite{Jewitt13}. See also figure 2 in this
  paper. Note that tail G is the youngest tail, and does not appear in
  the HST observations. The directions of celestial North and East are
  indicated, as 
  well as the direction to the Sun and the asteroid velocity vector.
\label{fig1}}
\end{figure}

\clearpage
\begin{figure}[ht]
  \begin{tabular}{@{}cc@{}}
    \includegraphics[angle=-90,width=.5\textwidth]{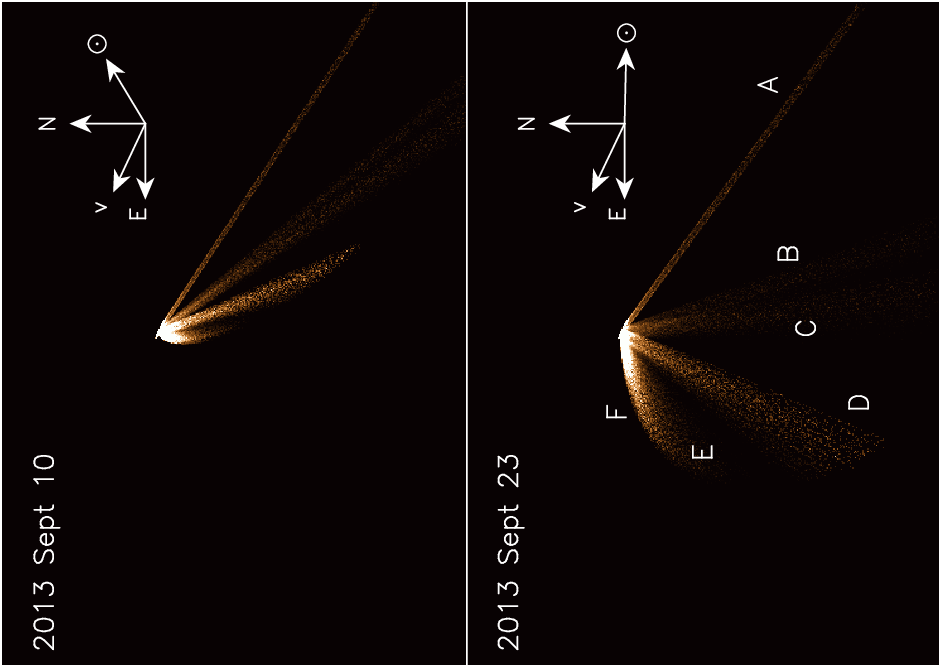} &
    \includegraphics[angle=-90,width=.5\textwidth]{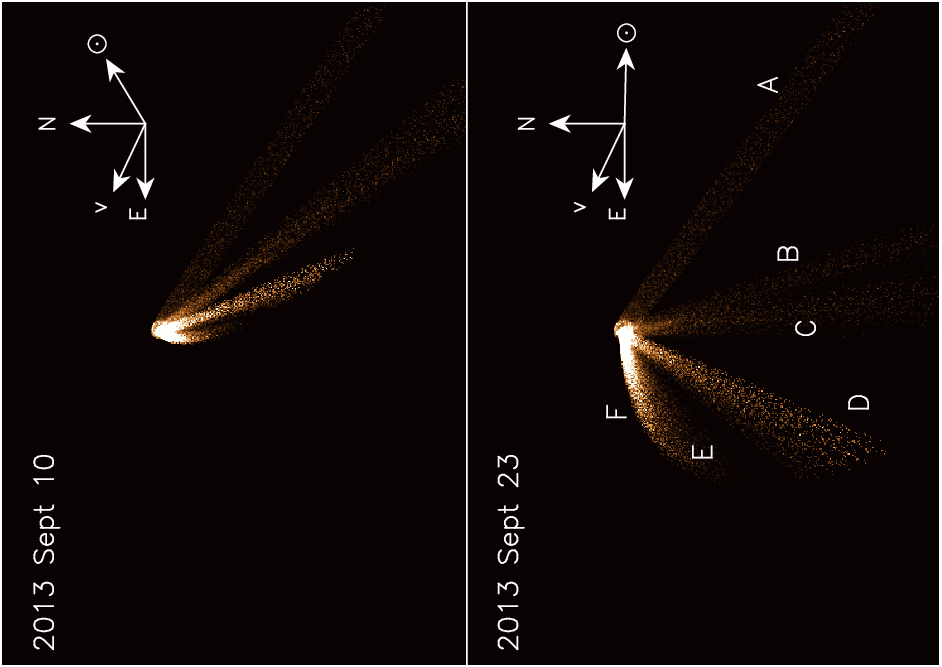} \\
   \end{tabular}
\caption{Model simulations of the Hubble Space Telescope images by
  \cite{Jewitt13} (see their figure 1), at two epochs. The left panels
  correspond to an anisotropic ejection model, where the particles are
  ejected from the equator of a rotating nucleus
  with spin axis perpendicular to the orbit plane. The right panels
  correspond to an isotropic ejection model, with the same input parameters
  as the anisotropic model. In the lower panels, each tail is 
  labeled according the nomenclature by \cite{Jewitt13}. The panels are
  23000 km in width, the same as in figure 1 by  \cite{Jewitt13}, to
  facilitate comparison.  The directions
  of celestial North and East are indicated, as 
  well as the direction to the Sun and the asteroid velocity
  vector. 
\label{fig2}}
\end{figure}

\clearpage
\begin{figure}[ht]
\centering
  \begin{tabular}{@{}cc@{}}
    \includegraphics[angle=-90,scale=0.46]{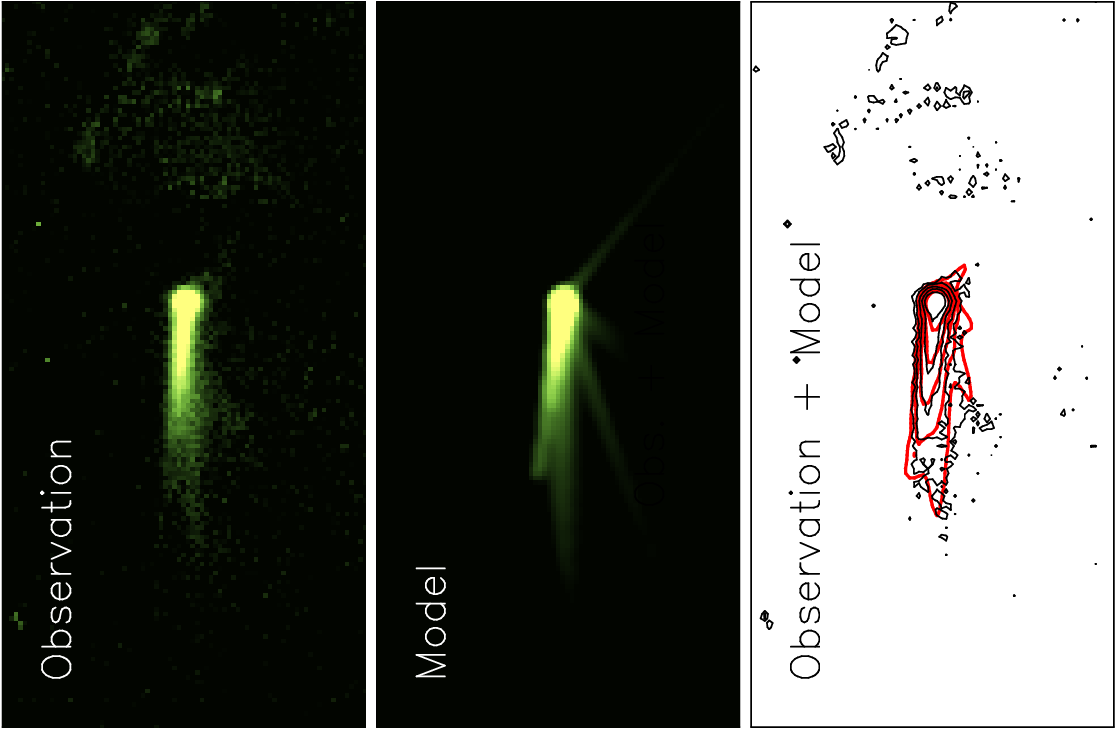} &
    \includegraphics[angle=-90,width=.5\textwidth]{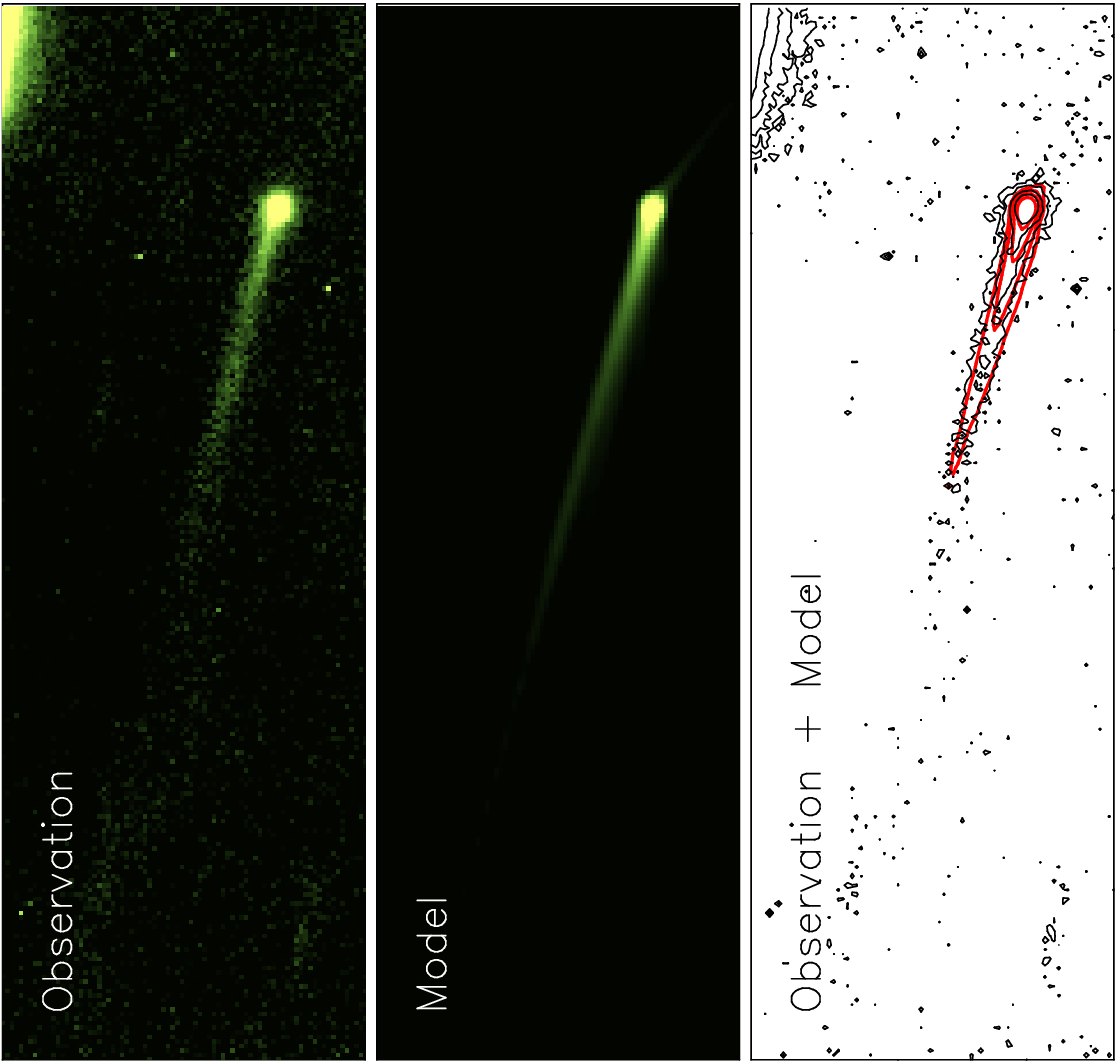} \\
   \end{tabular}
\caption{Left panels: observation and model
  simulation of the 2013 October 7 image. The lowermost panel show the comparison of
  the observed and modeled isophotes. The innermost isophote level is
  3.8$\times$10$^{-14}$ solar disk intensity units, and the isophotes
  decrease in a factor of 2 between consecutive levels. Right panels:
  observation and model 
  simulation of the 2013 November 8 image. The lowermost panel show the comparison of
  the observed and modeled isophotes. The innermost isophote level is
  2$\times$10$^{-14}$ solar disk intensity units, and the isophotes
  decrease in a factor of 2 between adjacent levels. The dimensions of
  the images are the same as in Figure 1.
\label{fig3}}
\end{figure}

\clearpage
\begin{figure}[ht]
\centerline{\includegraphics[scale=0.8,angle=-90]{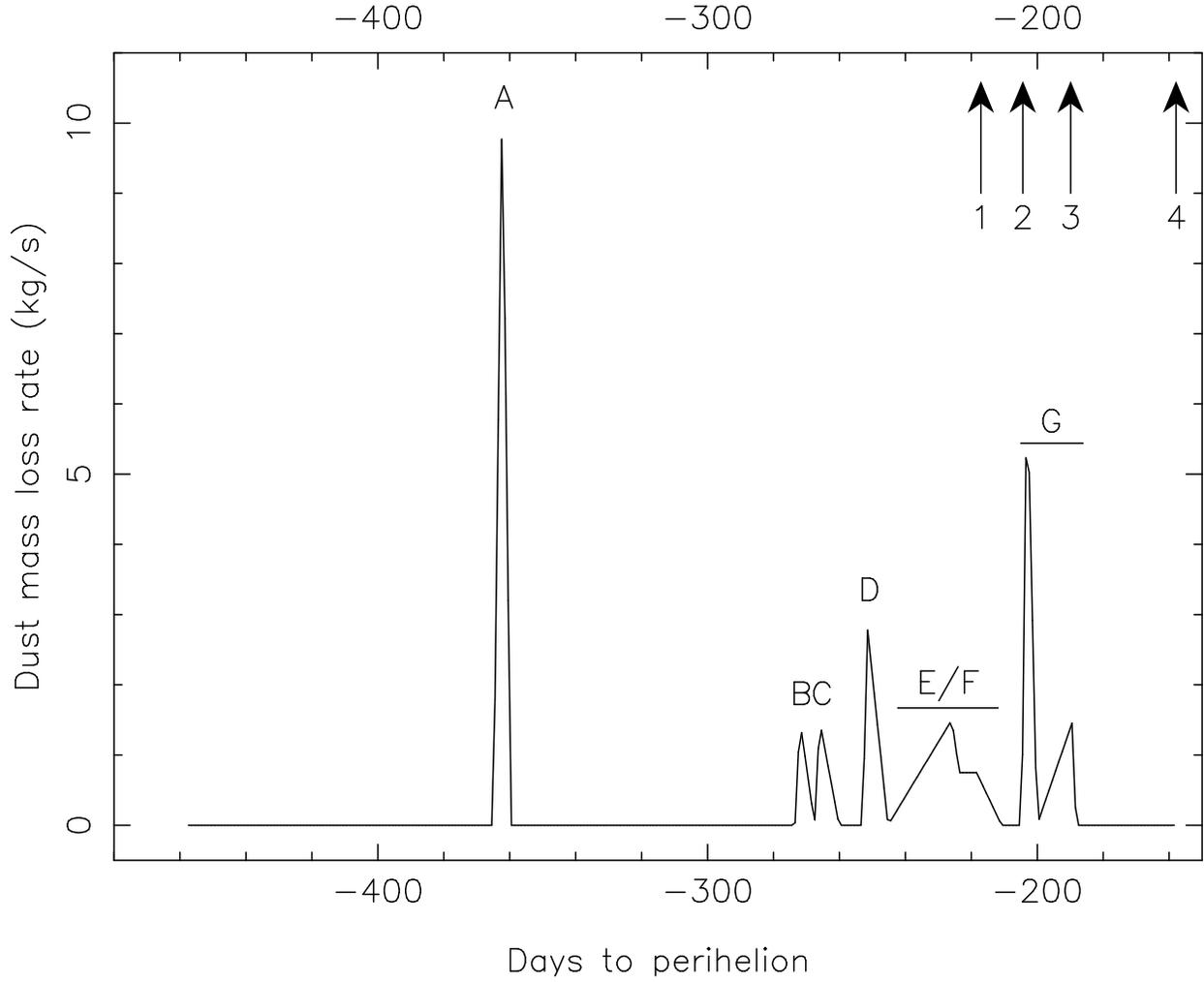}}
\caption{Modeled dust mass loss rate from P/2013 P5 as a function of
  time to perihelion. The sharp peaks of dust ejection are associated
  to  the tails labeled ``A'' to ``G'' (see Figures 2 and 3). The
  arrows indicate the observation dates of the HST (1 and 2), and
  the GTC (3 and 4) data.
\label{fig4}}
\end{figure}

\clearpage

\begin{deluxetable}{ccccccc}
\tablewidth{0pt}
\tablecaption{Log of the observations.}
\tablehead{
\colhead{Date (UT)} & \colhead{r$_h$(AU)} & \colhead{$\Delta$(AU)} & 
\colhead{$\alpha$($^\circ$)} & \colhead{$\delta$($^\circ$)} &  \colhead{$m$}
& \colhead{$H$} }
\startdata
2013 Oct.7 22:18 & 2.077 & 1.204 & 17.7 & --4.10 & $g^\prime$=21.3$\pm$0.1 &18.5$\pm$0.1 \\  
2013 Oct.7 22:26 & 2.077 & 1.204 & 17.7 & --4.10 & $r^\prime$=20.8$\pm$0.1 &18.0$\pm$0.1 \\
2013 Nov.8 21:04 & 2.038 & 1.462 & 27.0 & --2.68 &$r^\prime$=21.5$\pm$0.1&18.0$\pm$0.1 \\
\enddata
\end{deluxetable}

\begin{thebibliography}{}

\bibitem[Bertini(2011)]{Bertini11} Bertini, I. 2011, Planet. Space
  Sci., 59, 365

\bibitem[Bowell et al.(1989)]{Bowell89}Bowell,E., Hapke,B., Domingue,D., et
  al. 1989, in Asteroids II, University of Arizona Press, Tucson 

\bibitem[Cepa et al.(2000))]{Cepa00} Cepa, J., Aguiar, M., Escalera,
  V. et al. 2000, Poc. SPIE, 4008, 623

\bibitem[Cepa(2010))]{Cepa10} Cepa, J. 2010, Highlights of Spanish
  Astrophysics V, Astrophysics and Space Science Proceedings, 
 Springer-Verlag, p. 15

\bibitem[Cox(2000)]{Cox00}Cox, A. 2000, Allen's Astrophysical
  Quantities (4th ed., New York: Springer) 

\bibitem[Edoh(1983)]{Edoh83} Edoh, O. 1983, PhD thesis, Univ. Arizona

\bibitem[Fukugita et al.(1996)]{Fukugita96}Fukugita, M., 
Ichikawa, T., Gunn, J. E., et al. 1996, \aj 111, 1748

\bibitem[Fulle(1989)]{Fulle89} Fulle, M., 1989, A\&A, 217, 283

\bibitem[Fulle et al.(2010)]{Fulle10} Fulle, M., Colangeli, L.,
  Agarwal, J., et al. 2010, A\&A, 522, 63

\bibitem[Hsieh \& Jewitt(2006)]{Hsieh06} Hsieh, H.H., \& Jewitt,
  D. 2006, Science, 312, 561

\bibitem[Jewitt et al.(2010)]{Jewitt10}Jewitt, D., Weaver, H., 
Agarwal, J. et al. 2010 Nature, 467, 817

\bibitem[Jewitt(2012)]{Jewitt12} Jewitt, D. 2012 \aj, 143, 21

\bibitem[Jewitt et al.(2013)]{Jewitt13} Jewitt, D., Agarwal, J., 
  Weaver, H., et al. 2013 \apjl, 778, L21

\bibitem[Masiero et al.(2013)]{Masiero13}Masiero, J. R., 
Mainzer, A. K., Bauer, J. M., et al. 2013, \apj, 770, 7

\bibitem[Micheli et al.(2013)]{Micheli13}Micheli, M., Tholen, 
D. J., Primak, N., et al. 2013, Minor Planet Electronic Circulars, 37

\bibitem[Moreno et al.(2010)]{Moreno10} Moreno, F., Licandro, J.,
  G.-P. Tozzi, et al. 2010, \apj, 718, L132

\bibitem[Moreno et al.(2012a)]{Moreno12a} Moreno, F., Pozuelos, F.,
  Aceituno, F., et al. 2012a, \apj, 752, 136

\bibitem[Moreno et al.(2012b)]{Moreno12b} Moreno, F., Licandro, J., \&
  Cabrera-Lavers, A. 2012b, \apj, 761, L12

\bibitem[Moreno et al.(2013)]{Moreno13} Moreno, F., Licandro, J. \&
  Ortiz, J.L 2013, \apj, submitted

\bibitem[Pravec et al.(2002)]{Pravec2002} Pravec, P., Harris, A.W, \&
  Michalowski, T. 2002, in Asteroids III, University of Arizona Press 

\bibitem[Snodgrass et al.(2010)]{Snodgrass10}Snodgrass, C., Tubiana,
  C., Vincent, J-B. et al. 2010 Nature, 467, 814


\end{thebibliography}
\end{document}